\documentclass[showpacs,preprintnumbers,amsmath,amssymb]{revtex4}

\usepackage{graphicx}
\usepackage{dcolumn}
\usepackage{bm}
\begin{document}
\preprint{Nuclear Theory}
\title{Nuclear effects in Neutrino Nuclear Cross-sections}
\author{S.K.Singh and M. Sajjad Athar}
\affiliation{Department of Physics, Aligarh Muslim University, Aligarh-202 002, India.}
\date{\today}
 
\begin{abstract}
Nuclear effects in the quasielastic and inelastic scattering of neutrinos(antineutrinos) from nuclear targets have been studied. The calculations are done in the local density approximation which take into account the effect of nucleon motion as well as renormalisation of weak transition strengths in the nuclear medium. The inelastic reaction leading to production of pions is calculated in a $\Delta$ dominance model taking into account the renormalization of $\Delta$ properties in the nuclear medium.
\end{abstract}
\maketitle 

Recently (anti)neutrino nuclear cross sections are being measured on nuclear targets in connection with the neutrino oscillation experiments specially by K2K and MiniBooNE collaborations. These measurements are made in the energy region of neutrino energy peaking around 0.7GeV(MiniBooNE) and 1.3GeV(K2K). In this energy region, the most important processes contributing to the neutrino nuclear cross sections are the quasielastic scattering and inelastic scattering leading to pion production. In this energy region, the nuclear effects arising due to Fermi motion, binding energy and nucleon-nucleon correlations in the nuclear medium are important for the quasielastic reactions. In the case of inelastic reactions, where the pions are produced, the final state interaction of pions also has to be taken into account. In the following, we describe a model in which these effects are taken into account and present the numerical results. 

\vspace{5mm}
{\bf 1. Quasielastic Reaction}

The cross section for quasi-elastic charged lepton production is calculated in local density approximation\cite{singh} using electroweak nucleon form factors of Bradford et al.\cite{bradford} with axial dipole mass ${M}_{A}$=1.05GeV and vector dipole
mass ${M}_{V}$=0.84GeV. The Fermi motion and the Pauli blocking effects in
nuclei are included through the imaginary part of the Lindhard function
for the particle hole excitations in the nuclear medium. The
renormalization of the weak transition strengths are calculated in the random
phase approximation(RPA) through the interaction of the p-h excitations as
they propagate in the nuclear medium using a nucleon-nucleon potential
described by pion and rho exchanges. The effect of the Coulomb distortion
of muon in the field of final nucleus is taken into account
using a local version of the modified effective momentum
approximation. 

The total cross section $\sigma(E_\nu)$ for the charged current neutrino induced reaction on a nucleon inside the nucleus in a local Fermi gas model is written as\cite{prd1}:

\begin{eqnarray}
\sigma(E_\nu)=-\frac{2{G_F}^2\cos^2{\theta_c}}{\pi}\int^{r_{max}}_{r_{min}} r^2 dr \int^{p_\mu^{max}}_{p_\mu^{min}}{p_\mu}^2dp_\mu \int_{-1}^1d(cos\theta)
\frac{1}{E_{\nu_\mu} E_\mu} L_{\mu\nu}{J^{\mu\nu}} Im{U_N(q_0, {\bf q})}.
\end{eqnarray}
where $L_{\mu\nu}=\sum L_\mu {L_\nu}^\dagger$ and ${J^{\mu\nu}}={\bar\sum}\sum J^\mu {J^\nu}^\dagger$ 
\begin{eqnarray}
L_{\mu}&=&\bar{u}(k^\prime)\gamma_\mu(1-\gamma_5)u(k)\nonumber\\
J^\mu&=&\bar{u}(p^\prime)[F_{1}(q^2)\gamma^\mu + F_{2}(q^2)i{\sigma^{\mu\nu}}{\frac{q_\nu}{2M}}
 + F_{A}(q^2)\gamma^\mu\gamma_5 + F_{P}(q^2)q^\mu\gamma_5]u(p).\nonumber
\end{eqnarray}
$q(=k-k^\prime)$ is the four momentum transfer. $U_N$ is the Lindhard function for the particle hole excitation\cite{singh}. The form factors $F_1$, $F_2$, $F_A$ and $F_P$ are isovector electroweak form factors.

Inside the nucleus, the Q-value of the reaction and Coulomb distortion of outgoing lepton are taken into account by modifying the $Im{U_N(q_0, {\bf q})}$ by $Im{U_N(q_0-V_c(r)-Q, {\bf q})}$. Furthermore, the renormalization of weak transition strength in the nuclear medium in a random phase approximation(RPA) is taken into account by considering the propagation of particle hole(ph) as well as delta-hole($\Delta h$) excitations resulting into the modification of the various weak coupling strengths except the charge coupling. These considerations lead to a modified hadronic tensor $J^{\mu\nu}_{RPA}$ for which expressions are given in Refs.\cite{singh,prd1}. 
\begin{figure}
\caption{R=$\frac{\sigma^{A}}{A\sigma^{free}}$  vs Neutrino(Antineutrino) Energy}
\includegraphics[height=.3\textheight, width=1.0\textwidth]{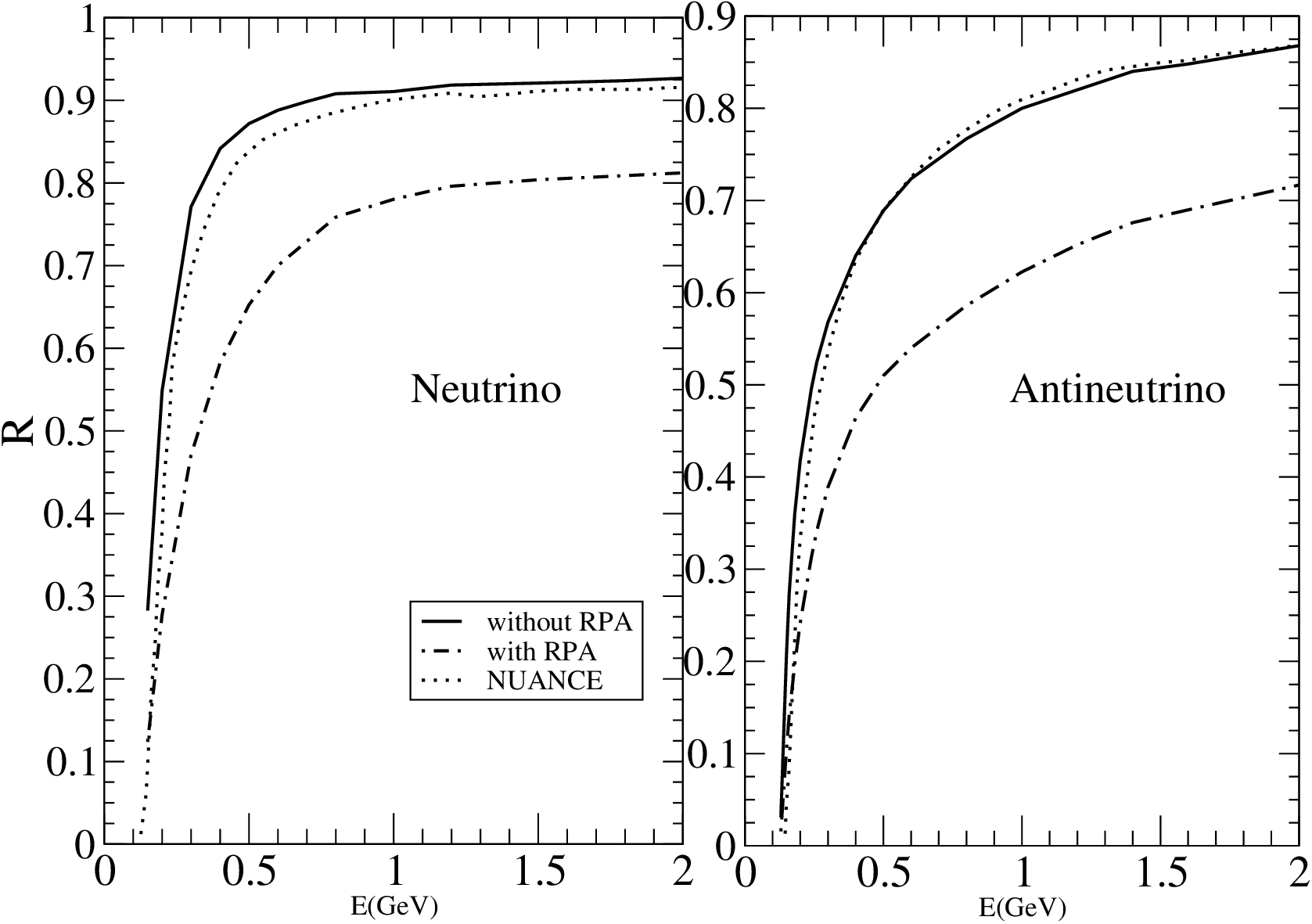}
\end{figure}

\vspace{5mm}
{\bf 2. Inelastic Reaction}

The cross sections for pion production is calculated using the $\Delta$ dominance model. In this model, the weak hadronic currents interacting with the nucleons in the nuclear medium excite a $\Delta$ resonance which decays into pions and nucleons. The pions interact with the nucleus inside the nuclear medium before coming out. The final state interaction of pions leading to elastic, charge exchange scattering and the absorption of pions lead to reduction of pion yield. The nuclear medium effects on $\Delta$ properties lead to modification in its mass and width which are taken from a theoretical model to explain elastic and pion induced pion production processes analysed in this model by Oset et al.~\cite{Oset}. Using a modified mass $M_{\Delta} \rightarrow M_{\Delta}+ {Re}\Sigma_{\Delta}$ and modified width $\Gamma_{\Delta} \rightarrow \tilde\Gamma_{\Delta} - 2{Im}\Sigma_{\Delta}$, where $\tilde\Gamma_{\Delta}$ is reduced width of $\Delta$ due to Pauli blocking of nucleons in the $\Delta \rightarrow {N} \pi$ decay and $\Sigma_{\Delta}$ is the self energy of $\Delta$ calculated in nuclear many body theory using local density approximation. 

The total scattering cross section for the neutrino induced charged current lepton production process in the nucleus in the local density approximation is given by\cite{prd1}
\begin{eqnarray}
\sigma&=&\frac{G_F^{2}cos^2{\theta_c}}{256\pi^3}\int \int {d{\bf r}}\frac{d\bf{k^\prime}}{E_k E_{k^\prime}}\frac{1}{MM_\Delta}\left[\rho_p({\bf r})+\frac{1}{3}\rho_n({\bf r})\right] \frac{\frac{\tilde\Gamma}{2}-Im\Sigma_\Delta}{(W- M_\Delta-Re\Sigma_\Delta)^2+(\frac{\tilde\Gamma}{2.}-Im\Sigma_\Delta)^2}
L_{\mu\nu}J^{\mu\nu}
\end{eqnarray}
where $s=(p+k)^2$, W is the $\Delta$ invariant mass, $ M(M_\Delta)$ is the nucleon(delta) rest mass, $\Gamma_\Delta$ is the $\Delta$ width, $L_{\mu\nu}$ is leptonic tensor and hadronic tensor $J^{\mu\nu}$ is defined in terms of the hadronic matrix element $J^\mu=\bar\Psi_\alpha A^{\alpha\mu} \Psi$, $\Psi_\alpha$ is Rarita Schwinger wave function for $\Delta$, $\Psi$ is the nucleon wave function, $A^{\alpha\mu}$ is the N-$\Delta$ transition vertex given in terms of $N-\Delta$ transition form factors which are taken from the work of Lalakulich et al.\cite{Lalakulich}. The modified width and mass of $\Delta$ are described in terms of the self energy $\Sigma$ as~\cite{Oset}:
\begin{eqnarray}
M_{\Delta} \rightarrow {\tilde M_{\Delta}}&=&M_{\Delta}+\mbox{Re}\Sigma_{\Delta}=
M_{\Delta} + 40 \frac{\rho}{\rho_{0}}MeV ~~and \nonumber\\
\frac{\Gamma}{2} &\rightarrow&  \frac{\tilde\Gamma}{2} - Im{{\Sigma}_{\Delta}}\nonumber\\
=\frac{\tilde\Gamma}{2} &+& [C_{Q}\left (\frac{\rho}{{\rho}_{0}}\right )^{\alpha}+C_{A2}\left (\frac{\rho}{{\rho}_{0}}\right )^{\beta}+C_{A3}\left (\frac{\rho}{{\rho}_{0}}\right )^{\gamma}]~~~~
\end{eqnarray}
where ${\tilde\Gamma}$ is Pauli corrected width. $C_{Q}$ accounts for the $\Delta N  \rightarrow
\pi N N$ process, $C_{A2}$ for the two-body absorption process $\Delta
N \rightarrow N N$ and $C_{A3}$ for the three-body absorption process $\Delta N N\rightarrow N N N$. The coefficients $C_{Q}$, $C_{A2}$, $C_{A3}$ and $\alpha$, $\beta$ and $\gamma$ are taken from Ref.~\cite{Oset}. We have taken energy dependent decay width for the $\Delta$.

The pions produced in this process are scattered and absorbed in the nuclear medium. This is treated in a Monte Carlo simulation using the  results of Vicente Vacas et al.\cite{Vicente} for the final state interaction of pions.

\vspace{5mm}
{\bf 3. Quasielastic like Inelastic Reactions}

In a nuclear medium when a neutrino or antineutrino interacts with a nucleon inside the nucleus, the $\Delta$ which is excited may disappear through two and three body absorption processes like $\Delta N
\rightarrow N N$ and $\Delta N N\rightarrow N N N$ and thus mimic a quasielastic reaction. These $\Delta$ absorption processes are described by the $C_{A2}$ and $C_{A3}$ terms in the expression of Im$\Sigma_\Delta$ given in Eq.4. Thus these terms alone taken in the numerator of Eq.(2) will describe the quasielastic like processes in which a $\Delta$ is excited and is reabsorbed leading to only leptons in the final state. These are labelled as quasielastic like inelastic reactions.
\begin{figure}[t]
\caption{$<\frac{d\sigma}{dQ^2}>$ vs $Q^2$ averaged over the MiniBooNE flux. The curves are normalised to unit area.}
\includegraphics[height=.3\textheight, width=1.0\textwidth]{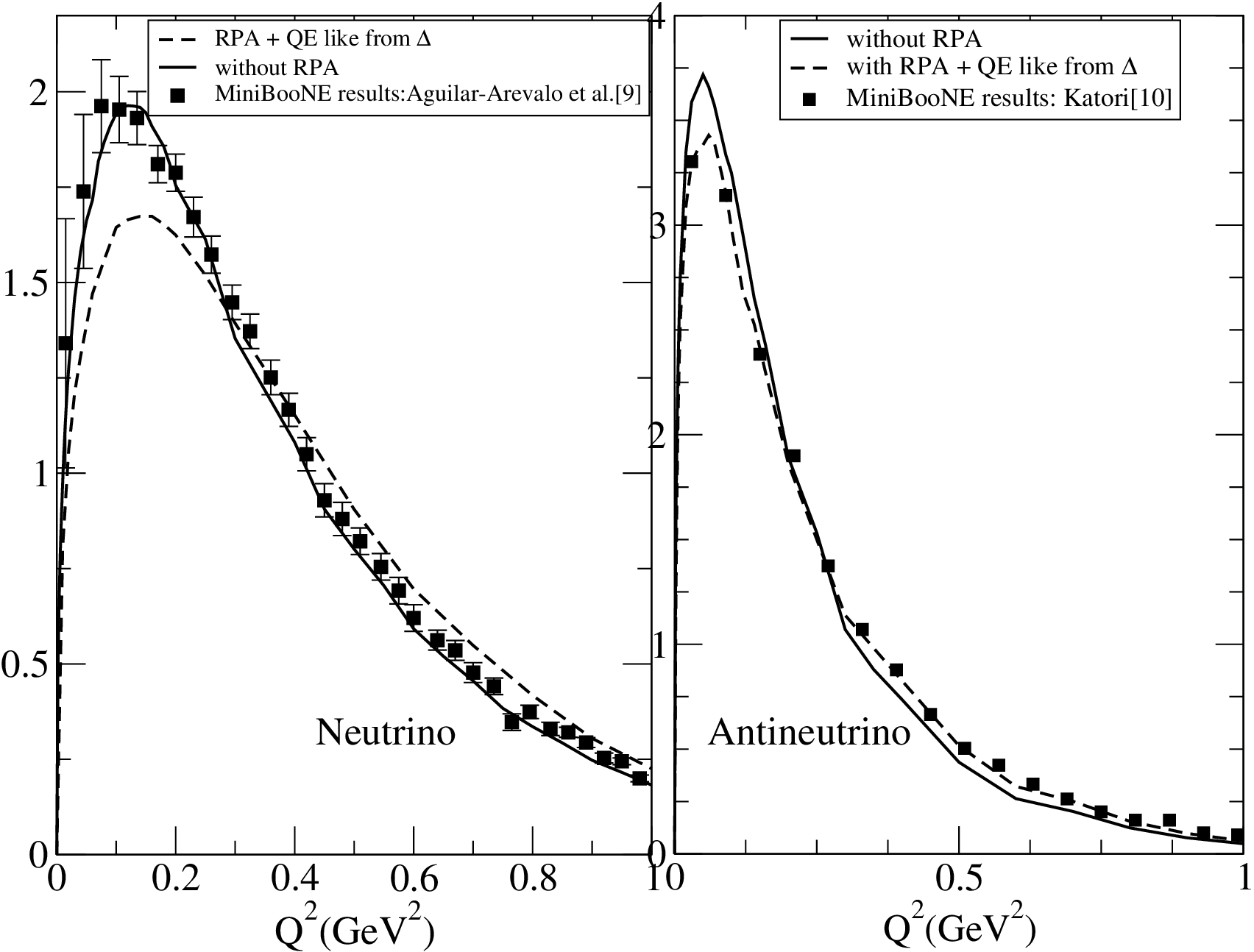}
\end{figure}
\begin{figure}
\caption{$\sigma^{cc1\pi^+}$  vs Neutrino Energy. The results shown with error bars are obtained by using the ratio $\frac{\sigma^{cc1\pi+}}{\sigma^{ccqe}}$ measured by the MiniBooNE collaboration\cite{boone1} and $\sigma^{ccqe}$ obtained in our model with RPA effects}
\includegraphics[height=.3\textheight, width=0.45\textwidth]{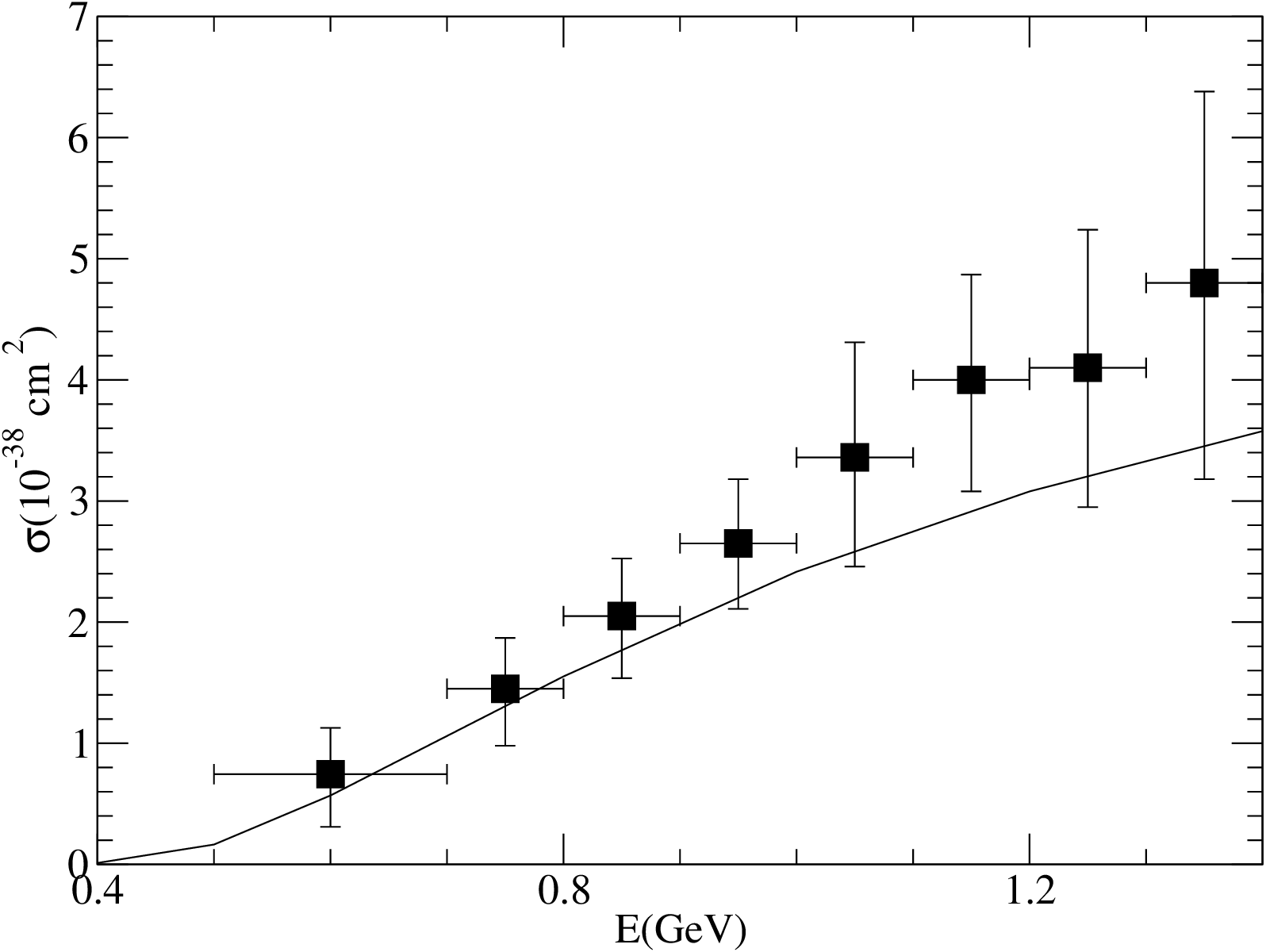}
\end{figure}

\vspace{5mm}
{\bf 4. Results and Discussion}

In Fig.1, we present the ratio of the charged current quasielastic lepton production cross section to the cross section on free nucleon i.e. $\frac{1}{N}\frac{\sigma(^{12}C)}{\sigma(free)}$ as a function of neutrino(Fig.1a) and antineutrino(Fig.1b) energies when $\sigma(^{12}C$ is calculated using Eq.(1). We find that with the incorporation of nuclear medium effects the reduction in the cross section is around $45\%$ at $E_{\nu}$=0.2GeV, $16\%$ at $E_{\nu}$=0.4GeV, $9\%$ at $E_{\nu}$=1.0GeV, $7\%$ at $E_{\nu}$=2GeV from the cross sections calculated for the free case. However, when we encorporate the RPA effects, the total reduction in the cross section is around $70\%$ at $E_{\nu}$=0.2GeV, $40\%$ at $E_{\nu}$=0.4GeV, $20\%$ at $E_{\nu}$=1.0GeV, $18\%$ at $E_{\nu}$=2GeV  from the cross sections calculated for the free case. These reductions have been found to be a bit more with antineutrinos induced process because of the some of the interference terms in the hadronic matrix element coming with opposite signs. The results have been compared with the results obtained in the Fermi gas model which has been used in the NUANCE Monte Carlo generator\cite{nuance} by the MiniBooNE collaboration\cite{boone1}. We find that the present results in the local Fermi gas model are similar to the results used in the NUANCE generator, but when RPA effects are used the cross sections are reduced.

In Fig.2, we have shown the results of the $Q^2$-distribution averaged over the MiniBooNE flux for $\nu_\mu$(Fig.4a) and ${\bar\nu}_\mu$(Fig.4b) flux. The results have been shown for the charged current quasielastic process in the local Fermi gas model with RPA effects as well as with the contribution from the charged current quasielastic-like process coming from the $\Delta$ disappearance. We find that in the peak region of $Q^2$ the contribution from the quasielastic-like process is 12-15\%. In these figures, we also show the preliminary results from MiniBooNE collaboration for $\nu_\mu$\cite{aguilar} and ${\bar\nu}_\mu$\cite{katori} respectively. We find that for the neutrino reactions the reduction in the differential cross section when calculated in the local Fermi gas model to the free case is around 30$\%$ in the peak region of $Q^2$. When RPA effects are also taken into account this reduction becomes 50$\%$ and decreases with the increase in $Q^2$. Similar results are obtained for antineutrino reactions.

In Fig.3, 
we have presented the results for CC1$\pi^+$ production cross section with nuclear medium and final state interaction effects. For one $\pi^+$ production process $\tilde\Gamma$ and $C_{Q}$ term in
$Im\Sigma_\Delta$ give contribution to the pion production. We find that the
nuclear medium effects lead to a reduction of around 12-15$\%$ for
neutrino energies $E_{\nu}$=0.6-3~GeV. When pion absorption
effects are also taken into account the total reduction in the cross section is around
$30-40\%$. In this figure, we have also shown the results of CC1$\pi^+$ production cross section obtained from the ratio of $\frac{\sigma^{cc1\pi+}}{\sigma^{ccqe}}$ vs $E_\nu$ measured by the MiniBooNE collaboration\cite{boone1} and $\sigma^{ccqe}$ obtained in our model with RPA effects. We find that our numerical results are in reasonable agreement with the experimentally reported results from MiniBooNE collaboration \cite{boone1}.

\end{document}